# Dark energy, co-evolution of massive black holes with galaxies, and ASTROD-GW*

Wei-Tou Ni[a,b]

[a]Shanghai United Center for Astrophysics (SUCA), Shanghai Normal University, 100 Guilin Road, Shanghai 200234, China

[b]Department of Physics, National Tsing Hua University, 101, Kuang Fu II road, Hsinchu, Taiwan, 300 ROC

e-mail: weitou@gmail.com; telephone: 86-15335151623; fax: 886-35723052

**Abstract**

The detection of low frequency band (100 nHz-100 mHz) and very low frequency band (300 pHz-100 nHz) gravitational waves (GWs) is important for exploration of the equation of state of dark energy and the co-evolution of massive black holes (MBHs) with galaxies. Most galaxies are believed to have a massive black hole in the galactic core. In the formation of these black holes, merging and accretion are the two main processes. Merging of massive black holes generate GWs which could be detected by space GW detectors and pulsar timing arrays (PTAs) to cosmological distances. LISA (Laser-Interferometric Space Antenna) is most sensitive to the frequency band 1 mHz-100 mHz, ASTROD-GW (ASTROD [Astrodynamical Space Test of Relativity using Optical Devices] optimized for Gravitational Wave detection) is most sensitive to the frequency band 100 nHz-1 mHz and PTAs are most sensitive to the frequency band 300 pHz-100 nHz. In this paper, we discuss the sensitivities and outlooks of detection of GWs from binary massive black holes in these frequency bands with an emphasis on ASTROD-GW. The GWs generated by the inspirals, merging and subsequent ringdowns of binary black holes are standard sirens to the cosmological distance. Using GW observations, we discuss the methods for determining the equation of state of dark energy and for testing the co-evolution models of massive black holes. ASTROD-GW is an optimization of ASTROD to focus on the goal of detection of GWs. The mission orbits of the 3 spacecraft forming a nearly equilateral triangular array are chosen to be near the Sun-Earth Lagrange points L3, L4 and L5. The 3 spacecraft range interferometrically with one another with arm length about 260 million kilometers. With 52 times longer in arm length compared to that of LISA, the strain detection sensitivity is 52 times better toward larger wavelength. The scientific aim is focused for gravitational wave detection at low frequency. The science goals include detection of GWs from MBHs, and Extreme-Mass-Ratio Black Hole Inspirals (EMRI), and using these observations to find the evolution of the equation of state of dark energy and to explore the co-evolution of massive black holes with galaxies.

# 1. Introduction

The existence of gravitational waves (GWs) is the direct consequence of general relativity and unavoidable consequences of all relativistic gravity theories with finite velocity of propagation (Einstein, 1916, 1918; Poincaré, 1905). Gravitational radiation is confirmed by binary pulsar orbit evolution (Weisberg and Taylor, 2004). The importance of gravitational-wave detection is twofold: (i) as probes to explore fundamental physics and cosmology, especially black hole physics and early cosmology, and (ii) as a tool in astronomy and astrophysics to study compact objects and to count them.

In this paper, we focus on using the GWs generated by massive black holes to explore the co-evolution of black holes with galaxies and to determine the dark energy equation of state. In section 2, we review classification of GWs briefly. In section 3, we present the current status and outlook of direct GW detection with an emphasis on the low frequency band (100 nHz-100 mHz) and very low frequency band (300 pHz-100 nHz), and point out that the ASTROD-GW has the best sensitivity in the frequency band 100 nHz-1 mHz). In section 4, we focus on ASTROD-GW. In section 5, we focus on GWs and GW background generated by BH co-evolution with galaxies. In section 6, we address to the issue of determination of the equation of state of dark energy by detection of individual GW sources generated by massive black holes. In section 7, we mention the scientific goal of space GW detectors to test general relativity. In section 8, we comment on the use of space GW detectors in probing the inflationary physics. In section 9, we present an outlook.

# 2. Classification of GWs

Similar to frequency classification of electromagnetic waves to radio wave, millimeter wave, infrared, optical, ultraviolet, X-ray and γ-ray etc., we can have the following complete Frequency Classification of Gravitational Waves (Figure 1) (Ni, 2010, 2008, 2005, 1997; Thorne, 1995):

(i) Ultra high frequency band (above 1 THz): Detection methods include Terahertz resonators, optical resonators, and ingenious methods to be invented.

(ii) Very high frequency band (100 kHz – 1 THz): Microwave resonator/wave guide detectors, optical interferometers and Gaussian beam detectors are sensitive to this band.

(iii) High frequency band (10 Hz – 100 kHz): Low-temperature resonators and laser-interferometric ground detectors are most sensitive to this band.

(iv) Middle frequency band (0.1 Hz – 10 Hz): Space interferometric detectors of short arm length (1000-100000 km) are most sensitive to this band.

(v) Low frequency band (100 nHz – 0.1 Hz): Laser-interferometer space detectors are most sensitive to this band.

(vi) Very low frequency band (300 pHz – 100 nHz): Pulsar timing observations are most sensitive to this band.

(vii) Ultra low frequency band (10 fHz – 300 pHz): Astrometry of quasar proper motions is most sensitive to this band.

(viii) Extremely low (Hubble) frequency band (1 aHz – 10 fHz): Cosmic microwave background experiments are most sensitive to this band.

(ix) Infra Hubble frequency band (below 1 aHz): Inflationary cosmological models give strengths of GWs in this band. They may be verified indirectly through the verifications of inflationary cosmological models.

For a summary of detection methods, please see Ni (2010).

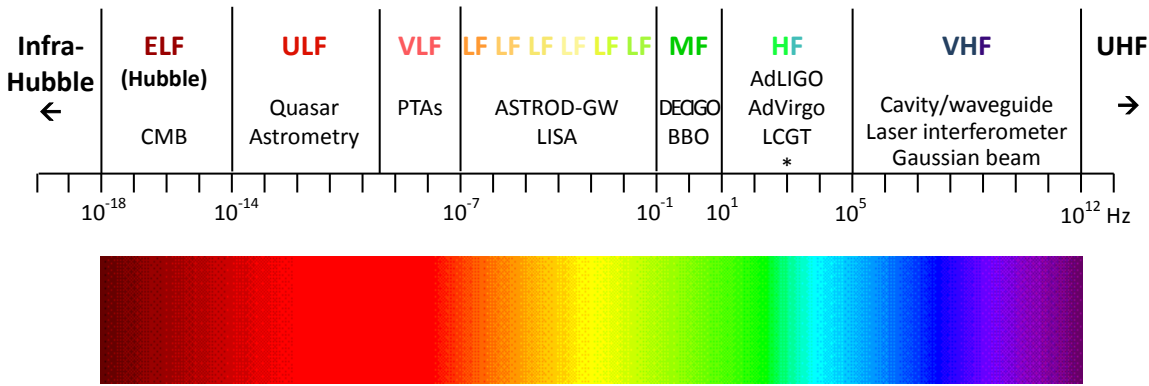

* AIGO, AURIGA, ET, EXPLORER, GEO, NAUTILUS, MiniGRAIL, Schenberg.

Figure 1. Frequency classification of GWs.

## 3. GW Detection

At present, the GW detection activities are focused on the high frequency band and the very low frequency band using ground interferometers and Pulsar Timing Arrays (PTAs) together with building experimental capacity of LISA (Laser Interferometric Space Antenna; LISA Study Team, 2000) for after 2020 launch to detect low frequency GWs. Ground interferometers already have enough sensitivity to set very interesting limits on the strengths of various GW sources. Analysis of the data from a LIGO two-year science run constrains the energy density $\Omega_{gw}$ of the stochastic gravitational-wave background normalized by the critical energy density of the Universe, around 100 Hz, to $6.9 \times 10^{-6}$ and improves on the indirect limit from the Big Bang nucleosynthesis at 100 Hz (The LIGO Scientific Collaboration and the Virgo Collaboration, 2009). Analysis of the data from the second science run of Virgo detector put direct upper limits on continuous gravitational wave emission from the Vela pulsar (The LIGO Scientific Collaboration and the Virgo Collaboration et al., 2011). The upper limits of about $2 \times 10^{-24}$ with 95% confidence have been obtained using three different methods, assuming that the gravitational wave emission follows the radio timing. These limits are below the indirect spin-down limit of $3.3 \times 10^{-24}$ for the Vela pulsar, defined by the energy loss rate inferred from observed decrease in Vela's spin frequency, and correspond to a limit on the star ellipticity of about $10^{-3}$. Advanced LIGO (http://www.advancedligo.mit.edu/summary.html),

Advanced Virgo (http://wwwcascina.virgo.infn.it/advirgo/) and LCGT (http://gw.icrr.u-tokyo.ac.jp/lcgt/) are currently under construction and will have the sensitivity to detect GWs from binary neutron-star mergers around 2016. LCGT is being constructed underground and will go cryogenic. ET (Einstein Telescope, http://www.et-gw.eu/) is under study for a 10-km underground cryogenic interferometer.

Several PTAs are actively searching for very low frequency GWs (Hobbs et al., 2010). Pulsar timing observations also become more and more interesting. For example, Jenet et al. (2004) ruled out the existence of a coalescing supermassive black-hole binary system (postulated in Sudou et al. (2003) from radio observations) at the centre of the radio galaxy 3C66B by using actual pulsar data from Kaspi et al. (1994).

In their white paper, Demorest et al. (2009) advocated that given current theoretical models, it is likely that very low frequency GWs will be detected by pulsar timing experiments within 5-10 yr. The first detections are expected to be of an isotropic, stochastic GW background created by coalescing supermassive binary black-hole systems. The comparison of current and planned GW detectors is shown in figure 2 (Demorest et al., 2009). We noticed that both PTAs and LISA lacks the sensitivity of detecting the MBH-MBH inspiral background in the frequency region 100 nHz – 10 μHz. ASTROD-GW (Ni, 2009a, 2010; Ni et al., 2009, 2010) fills this sensitivity gap as shown in figure 3. Moreover, the MBH (Massive Black Hole)-MBH GW backgrounds of all current models are above the ASTROD-GW sensitivity level.

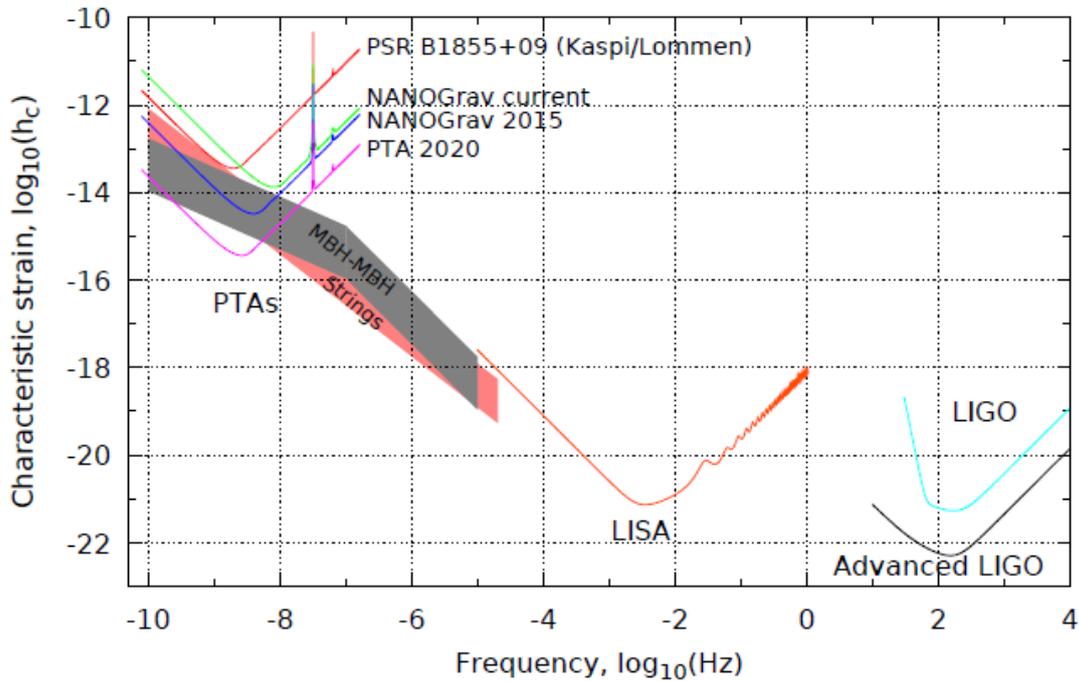

Figure 2: Comparison of current and planned GW detectors, showing characteristic strain ($h_c$) sensitivity versus frequency along with expected source strengths (Demorest et al., 2009). LIGO, LISA and PTAs occupy complementary parts of the GW spectrum. *There is an outstanding gap in the detection band 100 nHz to 10 μHz*. The gray strip is the region all current models of MBH-MBH GW backgrounds occupy.

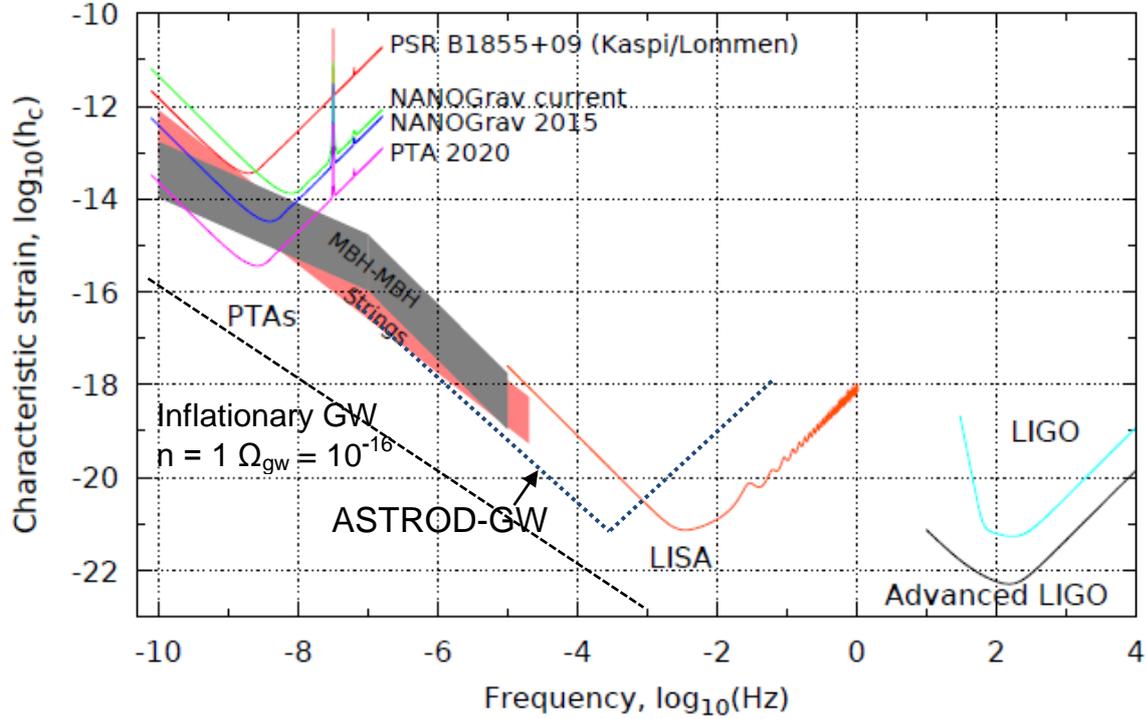

Figure 3: ASTROD-GW has the best sensitivity in the 100 nHz – 1 mHz band. *The outstanding gap in the detection band 100 nHz to 10 µHz is filled by ASTROD-GW. The MBH-MBH GW backgrounds of all current models are above the ASTROD-GW sensitivity level.* The dashed line in the bottom left corner corresponds to $\Omega_{gw} = 10^{-16}$ inflationary GW background ($\Omega_{gw}(f)$ in the figure is decadal density in terms of critical density of the universe defined to be $= (1/\rho_c)(d\rho_{gw}/d\log f)$ with $\rho_{gw}$ the energy density of the stochastic GW background and $\rho_c$ the present value of the critical density for closing the universe in general relativity.).

PTAs also give constraints on cosmic strings (Siemens et al., 2007). Depending on the string tension, the expected strain from cosmic strings could vary over many orders of magnitude. The band denoted by 'Strings' in Figure 2 and Figure 3 represents more optimistic predictions of cosmic string models (Demorest et al., 2009). With the ASTROD-GW sensitivity, these optimistic models could also be verified or constrained.

**4. ASTROD-GW**

ASTROD-GW (ASTROD [Astrodynamical Space Test of Relativity using Optical Devices] optimized for GW detection) is an optimization of ASTROD to focus on the goal of detection of GWs (Ni, 2009a, 2010; Ni et al., 2009, 2010). The scientific aim is focused for GW detection at low frequency. The mission orbits of the 3 spacecraft forming a nearly equilateral triangular array are chosen to be near the Sun-Earth Lagrange points L3, L4 and L5 (Figure 4).

The 3 spacecraft range interferometrically with one another with arm length about 260 million kilometers. After mission-orbit optimization (Ni et al., 2010; Wang and Ni, 2010, 2011, 2012; Men et al., 2009, 2010), the changes of arm length can be limited to ±1.732 × $10^{-4}$ AU or, fractionally, less than ±$10^{-4}$ in 20 years, and the Doppler velocities for the three spacecraft limited to about ±3 m/s. These values are well within the requirements of LISA and a number of technologies developed by LISA could be applied to ASTROD-GW. For the purpose of primordial GW detection (section 7), a 6-S/C formation for ASTROD-GW will be used for correlated detection of stochastic GWs.

Since the arm length is longer than LISA by 52 times, with 1-2 W laser power and LISA acceleration noise, the strain sensitivity of ASTROD-GW is 52 times lower than LISA in the lower frequency region, and is better than LISA and Pulsar Timing Arrays (PTPs) in the frequency band 100 nHz – 1 mHz. The instrumental noise density amplitude for ASTROD-GW is

$$S_{An}^{1/2}(f) = (1/L_A) \times \{[(1 + 0.5 (f/f_A)^2)] \times S_{Ap} + [1 + (10^{-4}/f)^2] (4S_a/(2\pi f)^4)\}^{1/2} \text{ Hz}^{-1/2}, \quad (1)$$

over the frequency range of 100 nHz < $f$ < 1 Hz with $f$ in unit of Hz. Here $L_A$ = 260 × $10^9$ m is the ASTROD-GW arm length, $f_A$ = c / (2π$L_A$) is the ASTROD-GW arm transfer frequency, $S_a$ = 9 × $10^{-30}$ $m^2$ $s^{-4}$ $Hz^{-1}$ is the white acceleration noise level, and $S_{Ap}$ = 10816 × $10^{-22}$ $m^2$ $Hz^{-1}$ is the (white) position noise level due to laser shot noise (Ni, 2009a, 2010; Ni et al., 2009, 2010). That for LISA is

$$S_{Ln}^{1/2}(f) = (1/L_L) \times \{[(1 + 0.5 (f/f_L)^2)] \times S_{Lp} + [1 + (10^{-4}/f)^2] (4S_a/(2\pi f)^4)\}^{1/2} \text{ Hz}^{-1/2}, \quad (2)$$

where $L_L$ = 5 × $10^9$ m is the LISA arm length, $f_L$ = c / (2π$L_L$) is the LISA arm transfer frequency, and $S_{Lp}$ = 4 × $10^{-22}$ $m^2$ $Hz^{-1}$ (Arun et al., 2009b). The noise power spectra curve for ASTROD-GW as compared with that of LISA is shown in Figure 5 (Ni et al., 2010; Ni, 2010). ASTROD-GW will complement LISA and PTAs in exploring single events and backgrounds of MBH-MBH binary GWs in the important frequency range 100 nHz - 1 mHz to study black hole co-evolution with galaxies (section 5) and dark energy issues (section 6).

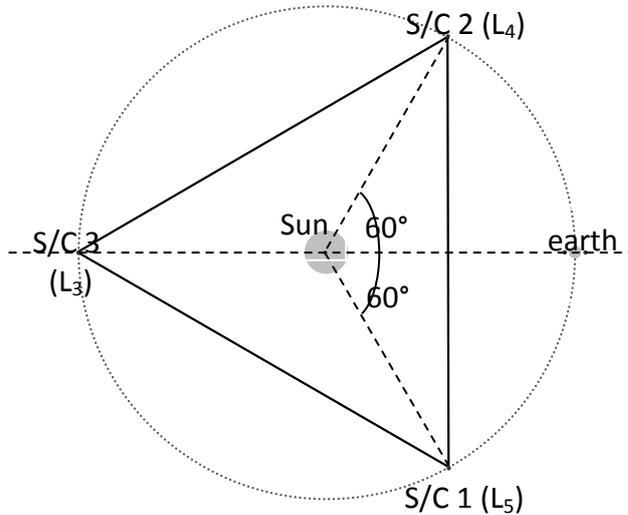

Figure 4: Schematic of ASTROD-GW mission orbit design.

Since the arm lengths among 3 spacecraft are changing, in order to attain the requisite sensitivity for ASTROD-GW, laser frequency noise must be suppressed below the secondary noises such as the optical path noise, acceleration noise etc. For suppressing laser frequency noise, we need to use time delay interferometry to match the time. We simulated the time delay numerically using CGC2.7 ephemeris framework (Wang and Ni, 2010, 2011, 2012a, 2012b). To conform to the ASTROD-GW planning, we worked out as an example a set of 20-year optimized mission orbits of ASTROD-GW spacecraft starting at June 21, 2028, and calculate the residual errors in the first and second generation time delay interferometry. For all the second generation time delay interferometry calculated, their interferometric optical path differences are less than 150 ns (50 m) which is well within the requirement for the ASTROD-GW mission.

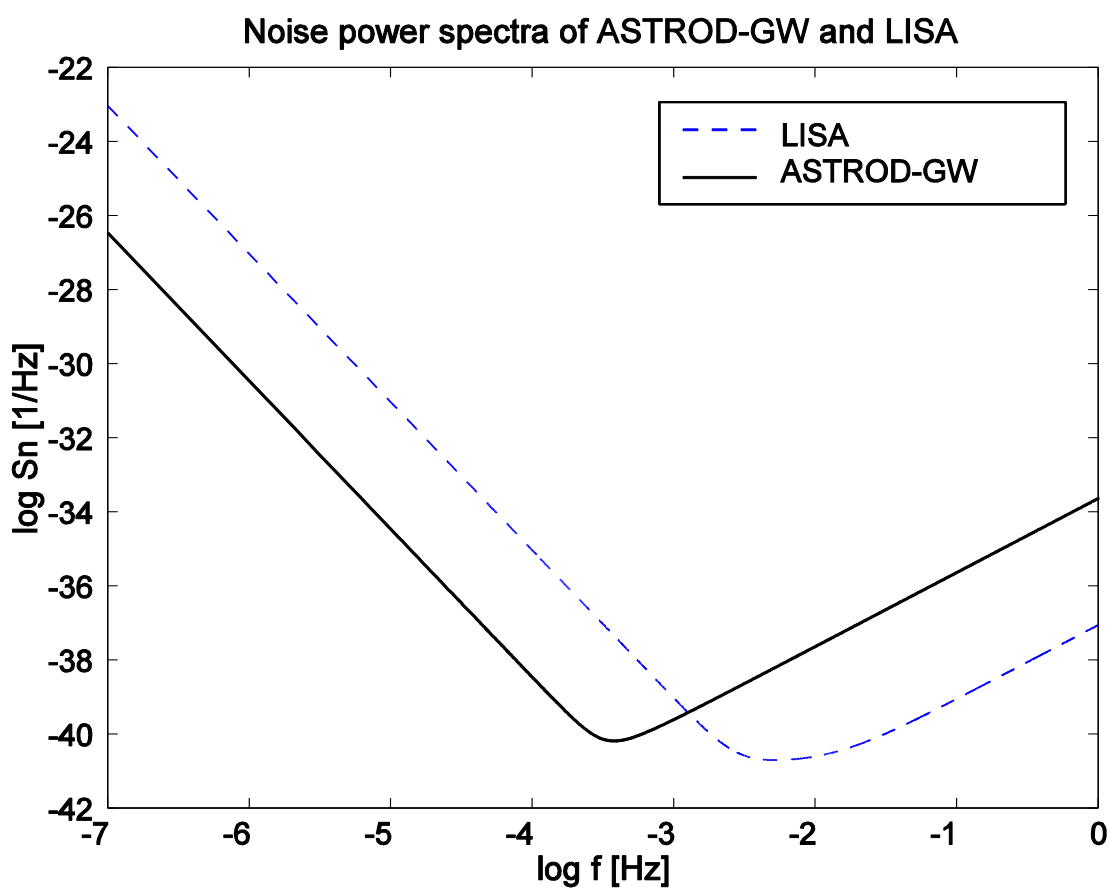

Figure 5. Strain noise power spectra of ASTROD-GW as compared with LISA.

The original ASTROD-GW orbits were chosen in the ecliptic plane (Men et al, 2010), the angular resolution in the sky has antipodal ambiguity and, near ecliptic poles, the resolution is poor, although over most of sky the resolution is good. We have revised the orbits of ASTROD-GW spacecraft to have small inclinations of 1-3° to the ecliptic plane to resolve this issue of antipodal degeneracy while keeping the variation of the arm lengths in the tolerable range. Now we describe this inclined orbit configuration with 3 S/C near $L_3$, $L_4$, and $L_5$ points.

A circular orbit of a spacecraft in the Newtonian gravitational central problem (one-body) in spherical coordinates (r, θ, φ) is:

$$r = a, \theta = 90°, \varphi = \omega t + \varphi_0, \qquad (3)$$

where a, ω, and $\varphi_0$ are constants. For spacecraft in this discussion, we have a = 1 AU, $\omega = 2\pi/T_0$ with $T_0 = 1$ sidereal year, and $\varphi_0$ is the initial phase in the coordinate considered. The spacecraft orbit at time t in Cartesian coordinates is

$$x = a \cos \varphi = a \cos(\omega t + \varphi_0); \; y = a \sin \varphi = a \sin(\omega t + \varphi_0); \; z = 0. \qquad (4)$$

We can transform this orbit actively into an orbit with inclination λ; and with the intersection of the orbit plane and the xy-plane (the cliptic plane) at the line $\varphi = \Phi_0$ in the xy-plane. The new spacecraft orbit is

$$\begin{pmatrix} x' \\ y' \\ z' \end{pmatrix} = \begin{pmatrix} a[1 - \sin^2\Phi_0(1 - \cos\lambda)]\cos\varphi + a \sin\Phi_0\cos\Phi_0(1 - \cos\lambda)\sin\varphi \\ a \cos\Phi_0\sin\Phi_0(1 - \cos\lambda)\cos\varphi + a[1 - \cos^2\Phi_0(1 - \cos\lambda)]\sin\varphi \\ -a \sin\Phi_0\sin\lambda\cos\varphi + a \cos\Phi_0\sin\lambda\sin\varphi \end{pmatrix} \qquad (5)$$

For the three orbits with inclination λ (in radian) of the three ASTROD-GW spacecraft --- S/C I, S/C II and S/C III, we choose:

$$\begin{aligned} &\text{S/C I: } \varphi_0(\text{I}) = 270°, \varphi_1(\text{I}) = 0°; \\ &\text{S/C II: } \varphi_0(\text{II}) = 150°, \varphi_1(\text{II}) = 120°; \\ &\text{S/C III: } \varphi_0(\text{III}) = 30°, \varphi_1(\text{III}) = 240°; \end{aligned} \qquad (6)$$

with S/C I near $L_5$ point, S/C II near $L_4$ point and S/C III near $L_3$ point. Defining the parameter ξ to be

$$\xi \equiv 1 - \cos \lambda = 0.5 \lambda^2 + O(\lambda^4), \qquad (7)$$

the arm vectors $\underline{V}_{\text{II-I}} \equiv \underline{r}^{\text{II}} - \underline{r}^{\text{I}}$, $\underline{V}_{\text{III-II}} \equiv \underline{r}^{\text{III}} - \underline{r}^{\text{II}}$ and $\underline{V}_{\text{I-III}} \equiv \underline{r}^{\text{III}} - \underline{r}^{\text{I}}$ are calculated to be:

$$\underline{V}_{\text{II-I}} = \begin{pmatrix} a[-(3/2)\cos\omega t - (3^{1/2}/2)\sin\omega t] + a\xi[(3^{1/2}/4)\sin\omega t + (3/4)\cos\omega t] \\ a[-(3/2)\sin\omega t + (3^{1/2}/2)\cos\omega t] + a\xi[(3/4)\sin\omega t - (3^{1/2}/4)\cos\omega t] \\ a \sin\lambda [(3^{1/2}/2)\sin\omega t - (3/2)\cos\omega t] \end{pmatrix}, \qquad (8)$$

$$\underline{V}_{\text{III-II}} = \begin{pmatrix} 3^{1/2} a \sin\omega t - (3^{1/2}/2) a \xi \sin\omega t \\ -3^{1/2} a \cos\omega t + (3^{1/2}/2) a \xi \cos\omega t \\ -3^{1/2} a \sin\lambda \sin\omega t \end{pmatrix}, \qquad (9)$$

$$\underline{V}_{\text{I-III}} = \begin{bmatrix} a[(3/2)\cos\omega t - (3^{1/2}/2)\sin\omega t] + a\,\xi[(3^{1/2}/4)\sin\omega t - (3/4)\cos\omega t] \\ a[(3/2)\sin\omega t + (3^{1/2}/2)\cos\omega t] + a\,\xi[-(3/4)\sin\omega t - (3^{1/2}/4)\cos\omega t] \\ a\sin\lambda\,[(3^{1/2}/2)\sin\omega t + (3/2)\cos\omega t] \end{bmatrix}, \qquad (10)$$

with arm lengths

$$|\underline{V}_{\text{II-I}}| = 3^{1/2}\,a\,[(1-\xi/2)^2 + \sin^2\lambda\,\sin^2(\omega t - 60°)]^{1/2},$$

$$|\underline{V}_{\text{III-II}}| = 3^{1/2}\,a\,[(1-\xi/2)^2 + \sin^2\lambda\,\sin^2(\omega t)]^{1/2},$$

$$|\underline{V}_{\text{I-III}}| = 3^{1/2}\,a\,[(1-\xi/2)^2 + \sin^2\lambda\,\sin^2(\omega t + 60°)]^{1/2}. \qquad (11)$$

The fractional arm length variation is within $(1/2)\sin^2\lambda$ which is about $10^{-4}$ for $\lambda$ about 1° and about $10^{-3}$ for $\lambda$ about 3° in the tolerable range.

The normalized unit vector $\underline{n}$ normal to the orbit configuration plane is:

$$\underline{n} = [\sin^2\lambda + (1-\xi/2)^2]^{1/2} \begin{bmatrix} -\sin\lambda\cos 2\omega t \\ -\sin\lambda\sin 2\omega t \\ (1-\xi/2) \end{bmatrix}. \qquad (12)$$

The geometric center $\underline{V}_c$ of the ASTROD-GW spacecraft configuration is

$$\underline{V}_c = \begin{bmatrix} -(1/2)\,\xi\,a\cos\omega t \\ (1/2)\,\xi\,a\sin\omega t \\ 0 \end{bmatrix}. \qquad (13)$$

There are 3 interferometers with 2 arms in the ASTROD-GW configuration. The geometric center of each of these 3 interferometers is at a distance of about 0.25 AU from the Sun. Numerical simulation and optimization of orbit configuration are underway using planetary ephemeris to take into account of the planetary perturbations. When LISA orbits around the Sun, it is equivalent to multiple detector arrays distributed in 1 AU orbit. The extension of ASTROD-GW is already of 1 AU. When ASTROD-GW orbits around the Sun, it is also equivalent to multiple detector arrays distributed in 1 AU orbit.

Now let us consider angular resolution of a coherent GW source. Consider first the LISA case as example. The detector formation of LISA is modulated in its orbit around the Sun. The azimuth modulation amplitude is $2\pi$ rad with inclination 1.05 rad (60°) so that the antenna pattern sweeps around the sky in one year. The antenna response is not isotropic but the averaged linear angular resolution (in a year) of monochromatic GW sources for LISA differs by less than a factor of 3 among all directions. *The angular resolution is basically proportional to the inverse of the strain signal to noise ratio.* If the inclination is of the order 0.03-0.1 rad (1-3°) for LISA, the polar resolution would be worsened by 30-90 times (approximately the ratio of 1.05 rad to 0.03-0.1 rad). The steradian localization in the celestial sphere is worsened by square of this factor; away from the polar region ($\theta \gg 0.03\text{-}0.1$ rad) by $\sin \theta$. If the signal to noise ratio is downgraded by 5 (as in eLISA/NGO in its low frequency part due to shorter arm length), the linear angular resolution is worsened by 5 times. ASTROD-GW has less sensitivity above 1 mHz compared with LISA, therefore the angular resolution will be worsened by both factors. In the 100 nHz-1 mHz region, ASTROD-GW has better sensitivity compared with LISA, in most part by 52 times. Hence, the angular resolution in the polar region is similar to that of LISA, while in other regions, the linear resolution is enhanced by roughly $52 \times \sin^{1/2} \theta$ (upgraded by 52 but downgraded by $\sin \theta$ in sterad [by $\sin^{1/2} \theta$ in rad]). Within a factor of 3, the averaged antenna pattern for ASTROD-GW away from the polar region is better by a factor of $52 \times \sin^{1/2} \theta$ compared to that of LISA. Since the antenna pattern of ASTROD-GW sweeps over the whole sky in half year as can been seen from Equation (12), the time of average needed is half a year instead of a year.

For more complicated sources like chirping GW sources from BBHs (binary black holes), one needs to do fitting in order to obtain the accuracy of the parameters. However, the tendency of accuracy of parameters, especially that for the luminosity distance is the same: for similar situations, it is proportional to the inverse of the strain signal to noise ration.

**5. BH Co-evolution with Galaxies**

There are observational evidences that massive black holes (MBHs) residing in most local galaxies. Relations have also been discovered between the MBH mass and the mass of host galaxy bulge, and between the MBH mass and the velocity-dispersion. These relations indicate that central MBHs are linked to the evolution of galactic structure (See, e.g., Volonteri, 2010b). Newly fueled quasar may come from the gas-rich major merger of two massive galaxies. Recent astrophysical evidences linked together these major galaxy mergers and the growth of supermassive black holes in quasars (Treister et al., 2010; Primack 2010). Treister et al. (2010) presented observations of the ratio of heavily obscured to unobscured quasars as a function of cosmic epoch up to z ≈ 3 and showed that a simple physical model describing mergers of massive, gas-rich galaxies matches these observations. In the context of this model, every obscured and unobscured quasar represents two distinct phases that result from a massive galaxy merger event. Much of the mass growth of the MBH occurs during the heavily obscured phase. The accretion time is determined to be around 100 million years. These observations provide additional evidence for a causal link between gas-rich galaxy mergers, and accretion onto the nuclear MBH.

Distant quasar observations indicate that MBH of billions of solar masses already existed less than a billion years after the Big Bang. At present, there are different theoretical proposals for scenarios of the initial conditions and formations of black holes. These scenarios include BH seeds from inflationary Universe and/or from the collapse of Population III stars, different accretion models and binary formation rates. All of these models generate MBH merging scenarios in galaxy co-evolution with GW radiations. Measurement of amplitude and spectrum of these GWs will tell the MBH cosmic history.

As we have just discussed, galaxy formation and evolution are closely related to the cosmological evolution of massive black hole formation. The standard theory of massive black hole formation is the merger-tree theory. In the merger-tree history, various Massive Black Hole Binary (MBHB) inspirals are acting. The GWs from these MBHB inspirals can be detected and explored to cosmological distances using space GW detectors. Although there are different merger-tree models and models with BH seeds, they all give significant detection rates for space GW detectors and Pulsar Timing Arrays (PTAs) (Sesana et al. 2008, Sesana et al., 2009, Demorest, et al. 2009). Gravitational wave (GW) observation in the 300 pHz - 1 Hz frequency band will be a major observation tool to study the co-evolution of galaxy with BHs. This frequency band covers the low frequency band (100 nHz - 100 mHz) and very low frequency band (300 pHz-100 nHz) GWs and is in the detection range of PTAs, LISA and ASTROD-GW. PTAs are most sensitive in the frequency range 300 pHz -100nHz, LISA space GW detector is most sensitive in the frequency range 1 mHz - 1 Hz, while ASTROD-GW is most sensitive in the frequency range 100 nHz -1 mHz (figure 3). PTAs have already been collecting data for detection of stochastic GW background from MBH binary mergers, and are aiming at detection around 2020 (Demorest, 2009). LISA and ASTROD-GW will be able to directly observe how massive black holes form, grow, and interact over the entire history of galaxy formation. These observations are significant and important to the study of co-evolution of galaxies with MBHs.

Recent studies (Sesana et al., 2008, 2009) of both stochastic GW background and resolvable source from massive black hole binary systems during galaxy co-evolution indicate that they would be detectable with PTAs and LISA. *With better sensitivity in the frequency bandwidth 100 nHz - 1 mHz, ASTROD-GW is in a unique position for this goal*. Together with PTAs and LISA, ASTROD-GW will be able to study this issue in detail.

At present, there are different theoretical scenarios for the initial conditions and formations of black holes. In one approach (Rubin et al., 2001; Khlopov et al., 2005; Dokuchaev et al., 2009), inflationary universe induces primordial massive BH clouds co-evolving with galaxies. Direct formation of supermassive black hole via multi-scale gas inflows in galaxy mergers is also a possibility. Mayer et al. (2010) reported simulations showing that mergers between massive protogalaxies naturally produce the conditions for direct collapse into a supermassive black hole with no need to suppress cooling and star formation (See also, Volonteri, 2010a). The mass range and maximum mass of Population III stars is also a relevant issue for seed BHs. A band in Figure 3 covers possible scenarios. ASTROD-GW will be able to detect GWs for every model in this region and to distinguish various scenarios for finding the history of BH and galaxy co-evolution.

With the detection of MBH merger events and background, the properties and distribution of MBHs could be deduced and underlying population models could be tested. Sesana et al. (2010) consider and compare ten specific models of massive black hole formation. These models are chosen to probe four important and largely unconstrained aspects of input physics used in the structure formation simulations, i.e., seed formation, metallicity feedback, accretion efficiency and accretion geometry. With Bayesian analyses to recover posterior probability distribution, they show that LISA has enormous potential to probe the underlying physics of structure formation. With better sensitivity in the frequency range 100 nHz - 1 mHz, ASTROD-GW will be able to probe the underlying physics of structure formation further. With the detection of the GW background of the MBH mergers, PTAs and ASTROD-GW will add to our understanding of the structure formation. Further studies along these lines are encouraged.

## 6. Determination of the Equation of State of Dark Energy

In the dark energy issue (Copeland et al., 2006), it is important to determine the value of $w$ in the equation of state of dark energy,

$$w = p/\rho, \tag{14}$$

for different epochs where $p$ is the pressure and $\rho$ the density of dark energy. For cosmological constant as dark energy, $w = -1$. From cosmological observations, our universe is close to being flat. In a flat Friedman Lemaître-Robertson-Walker (FLRW) universe, the luminosity distance is given by

$$d_L(z) = (1+z) \int_{0 \to z} (H_0)^{-1} [\Omega_m(1+z')^3 + \Omega_{DE}(1+z')^{3(1+w)}]^{-(1/2)} dz', \tag{15}$$

where $H_0$ is Hubble constant, $\Omega_{DE}$ is the present dark energy density parameter, and the equation of state of the dark energy $w$ is assumed to be constant. In the case of non-constant $w$ and non-flat FLRW universe, similar but more complicated expression can be derived. Here we show (15) for illustrative purpose. From the observed relation of luminosity distance vs. redshift $z$, the equation of state $w$ as a function of redshift $z$ can be solved for and compared with various cosmological models. Dark energy cosmological models can be tested this way. Luminosity distance from supernova observations and from gamma ray burst observations vs. redshift observations are the focus for the current dark energy probes. In the following we will address to the issue of using space GW detectors as dark energy probes.

In the merging of MBHs during the galaxy co-evolution processes, gravitational waveforms generated give precise, gravitationally calibrated luminosity distances to high redshift. The inspiral signals of these binaries can serve as standard candles/sirens (Schutz, 1986; Sathyaprakash and Schutz, 2009). Schutz (1986) first proposed to use GW observations of binaries to determine the Hubble constant. Finn and Chernoff (1993) analyzed the argument in more detail. For a nearby source with cosmological redshift negligible, the measured waveform $h$ has the following form:

$$h = [GM(m_i)/(c^2 r)] A(t) \cos[(t; m_i, \mathbf{S}_i)], \tag{16}$$

where $h$ represents typical component of gravitational wave strain, $m_i$'s are the binary's masses, $\mathbf{S}_i$'s are the coresponding spins, $M(m_i)$ is a function of the masses with dimension mass and $A(t)$ is a slowly varying dimensionless function dependent most strongly on parameters such as the source inclination (Hughes, 2009). For quadrupole wave form, $M(m_i)$ is the chirp mass

$$M = (m_1 m_2)^{3/5}/(m_1 + m_2)^{1/5}, \tag{17}$$

and A(*t*) is

$$A(t) \approx \pi(GMf_{GW}/c^3)^{2/3}, \tag{18}$$

with the chirp time $t_{\text{chirp}}$

$$t_{\text{chirp}} = (5/96)(GM/c^3)^{-5/3}(f_{\text{orbit}})^{-8/3}, \tag{19}$$

where $f_{\text{orbit}}$ is the orbit frequency of the binary. For cosmological binaries, all timescales that characterize the sources will be redshifted and equation (16) becomes

$$h = [G(1 + z)M(m_i)/(c^2 d_L)] A(t) \cos [(t; (1 + z)m_i, (1 + z)^2 S_i)], \tag{20}$$

where $d_L$ is the luminosity distance. That is, a binary with masses $m_i$ and spins $S_i$ at redshift $z$ has a phase evolution that looks just like a binary with masses $(1 + z)m_i$ and spins $(1 + z)^2 S_i$ locally.

Observations of massive black hole coalescences at cosmological distances by space-based GW detectors facilitate an accurate determination of the luminosity distance to the source. As an example, for quadrupole inspirals, basically the luminosity distance is given by amplitude, chirp time and orbital frequency as follows:

$$\text{Luminosity distance} \approx \text{velocity of light} \times \text{frequency}^{-2} \times t_{\text{chirp}}^{-1} \times \text{amplitude}^{-1}. \tag{21}$$

This well-known formula can be seen by using equations (7)-(20). The amplitude, frequency, and chirp rate of the binary can be measured from GW observation, and the luminosity distance can be inferred. The redshift z cannot be inferred from the observed signal for a signal with a redshift z and a chirp mass $M$ looks identical to a signal with no redshift and a chirp mass of M/(1 + z) (See, e.g., Hughes, 2009; and references therein).

The ground-based GW detectors would detect the inspirals and coalescence of stellar mass binaries within several hundred megaparsecs, while the space-based GW detectors like LISA and ASTROD-GW would study MBHs throughout the universe. The sensitivities of the space detector LISA and ASTROD-GW are shown in Figure 3 and Figure 5. The determination of distance using GW detection of binaries does not depends on distance ladder, and is depending only on the law of general relativity or other underlying gravitation theory. This is the way to measure the luminosity distance using gravitational wave (GW) signals which is different from the usual measurement using electromagnetic signals. A single detection with corresponding redshift measurement of associated electromagnetic counterpart would be useful to constrain dark energy model through Equation (15).

For LISA, the accuracy of luminosity distance determination for MBH-MBH mergers is expected to be 1-2 % for redshifts z < 3, degrading to ≈ 5 % for z ≈ 5 (Kocsis et al., 2007; Lang and Hughes, 2008; Arun et al., 2009a; Bloom et al., 2009). To use the luminosity distances for determining the equation of state for dark energy (Eq. (14)), one has to obtain redshifts of the host galaxies. High signal to noise ratio gives high angular resolution which facilitates the determination of optical association and redshift. Arun et al. (2007) showed that when higher signal harmonics are included in assessing the parameter estimation problem, the angular resolution increases by more than a factor of 10, making it possible for LISA to identify the host galaxy/galaxy cluster. LISA as a dark energy probe will be able to determine the equation-of-state parameter to a couple of per cents for $z < 1$ with limitation mainly comes from weak lensing (Arun et al., 2009a).

Since, weak lensing is a major limiting factor, let us address to the issue of weak lensing and delensing. Gravitational lensing randomly magnifies/de-magnifies electromagnetic and GW signals and degrades luminosity distance measurement. As investigated in Kocsis et al. (2006), at z=1, the weak lensing error in determining the redshift is about 2-3 %; at z=2, the weak lensing error is about 10 %; at z=3, the weak lensing error is about 20 %; at z=4, the weak lensing error is about 30 %. Various methods to reduce weak lensing uncertainties have been studied. Gunnarsson et al. (2006) used the observed properties of the foreground galaxies along the line of sight to the source to delense and reduce the dispersion due to lensing for source at z = 1.5 from about 7% to < 3%. Shapiro et al. (2010) proposed to use mapping shear and flexion of galaxy images to reduce the lensing error. They estimated that delensing with a 2D mosaic image from an Extremely Large Telescope could reduce distance errors by about 25–30 per cent for an MBHB at z=2. Including an additional wide shear map from a space survey telescope could reduce distance errors by nearly a factor of 2. Saini et al. (2010) proposed to use self-calibration to reduce systematic uncertainties in determining distance-redshift relation via gravitational radiation from merging binaries.

Van Den Broeck et al. (2010) in their study of weak lensing effects in the measurement of the dark energy equation of state with LISA presented up to 15 cases of different binary masses for each redshift at z=0.55, z=0.7 and z=1. These 15 cases include all possible combinations of observed component masses $m_{1,2}$ = {3.6 × $10^5$, 1.2 × $10^6$, 3.6 × $10^6$, 1.2 × $10^7$, 3.6 × $10^7$}$M_{sun}$. Without including error due to weak lensing, they obtained results consistent with the joint results of the LISA PE (Parameter Estimation) Taskforce (Arun et al., 2009b). Among other things, they noted that when the total mass is high, the termination frequency of the dominant harmonic will be low and the signal will have less power in LISA's frequency band, being "visible" only during the last orbits before merger. Both the low observed SNR and the short time over which the signal is observed will lead to relatively poor parameter estimation. This happened for individual binary mass to be around $10^7$ $M_{sun}$, and the identification of electromagnetic association is difficult. ASTROD-GW with SNR higher by 52 times and frequency band extended to 100 nHz will be suitable for a good observation of these cases and identification of their electromagnetic counterparts. Van Den Broeck et al. (2010) assumed an error at the level of 3-5% without delensing for z= 0.55-0.7 consistent with Kocsis et al. (2006) and emphasized that the limiting factor for being able to find the electromagnetic counterparts by association is the weak lensing error. They stressed on ways to reduce this error using astrophysical observation near line of sight.

For ASTROD-GW with 52 times less strain noise in most of the frequency band compared to LISA, the angular resolution and the localization in the sky of GW source from more massive MBHB is better. Since weak lensing affects both GW and electromagnetic propagation in the same way, the identification of electromagnetic association and redshift measurement should be readily applicable (Misra et al., 2012). These luminosity distance and redshift measurements may be possible to GW sources up to $z = 20$. Luminosity distance together with redshift measurement determines the equation of state of dark energy (Eq. (14)). The main limitation comes from weak lensing errors. Self-calibration methods of Saini et al. (2010) can be applied. With more studies and observations, the limitation from weak lensing will be clearer and hopefully suppressed to certain extent.

The determination of dark energy equation tests various dark energy models. We illustrate this with a model of the quintessential inflation. In this model, both inflation and dark energy (quintessence) are described by the same scalar field (Sahni et al. 2001; Sahni and Sami, 2004). In this scenario, during the radiative scheme, the scalar field equation of state locks to $w(\phi) \approx -1$ and behaves like an effective cosmological constant; during the current epoch of accelerated expansion $w(\phi) \leq -0.9$ and could be 5-10 % different from -1. A couple of percent accuracy in the dark energy equation would be able to distinguish this model from the cosmological constant model. This scenario on dark energy together with other scenarios could be empirically tested in the LISA and ASTROD-GW mission. As another example, we work on the Yang-Mills dark energy models. To distinguish the Yang-Mills dark energy model, high redshift observations are needed. Currently, only low redshift data have been accumulated with an upper limit up to z ~ 2. With LISA and ASTROD-GW, observations may be able extended to z ~ 3-20, and may enable various Yang-Mills dark energy models to be distinguished (Tong et al., 2012).

## 7. Comments on Probing the Inflationary Physics

CMB has imprints from inflationary GWs. The sensitivity of PLANCK Surveyor (www.rssd.esa.int/index.php?project=PLANCK) launched in 2009 has more than one order of magnitude improvement over WMAP. Dedicated CMB polarization observers (de Bernardis et al., 2009; Bock et al., 2008; KEK, 2008) would improve the sensitivity further by one order-of magnitude. This development would put the measurable tensor-to-scalar-ratio from inflation in the range of $10^{-3}$ corresponding to inflation energy scale of about $10^{15}$-$10^{16}$ GeV. This corresponds to $10^{-16}$-$10^{-17}$ in the predicted primordial $\Omega_{gw}$ for a number of simple slow roll inflation models (Smith et al., 2008).

The straight line in the bottom left corner of Figure 3 corresponds to $\Omega_{gw} = 10^{-16}$ inflationary GW background. For ASTROD-GW sensitivity to reach this region, we need (i) cross-correlated detection with multiple sets of detectors and (ii) further improvement of inertial sensor/accelerometer sensitivity. For optimally filtered Cosmic GW Background signal in cross-correlated detection, an enhancement factor of $(2Tf)^{1/4}$ can be obtained (Corbin and Cornish, 2006; Seto, 2006; Kawamura et al., 2006). For frequency in the band 100 nHz to 1 mHz and integration of 5 years, this enhancement factor is 2.3 to 23. With this enhanced sensitivity, it is barely enough to reach the $\Omega_{gw} = 10^{-16}$ line when a 6-S/C ASTROD-GW formation is used for correlated detection of stochastic GWs. To ensure the detection sensitivity at $\Omega_{gw} =$ (0.1-1) × $10^{-16}$ level, an improvement of one order of magnitude is needed in reducing the inertial sensor/accelerometer noise in the lower frequency region. After LISA Pathfinder and LISA, this seems to be a workable goal. However, the anticipated MBH-MBH GW background is above the 3-S/C ASTROD-GW sensitivity. If this background is detected, then the detectability of inflationary gravitational wave of the strength $\Omega_{gw} = 10^{-16}$-$10^{-17}$ from 6-S/C formation in the ASTROD-GW frequency region depends on whether this MBH-MBH GW 'foreground' could be separated due to different frequency dependence or other signatures. This is an important issue to be addressed.

For direct detection of primordial (inflationary, relic) GWs in space, one may go to frequencies lower or higher than LISA bandwidth (Ni, 2009b), where there are potentially less foreground astrophysical sources (Farmer and Phinney, 2004) to mask detection. DECIGO (Kawamura et al., 2006) and Big Bang Observer (Crowder and Cornish, 2005) look for GWs in the higher frequency range while ASTROD-GW looks for GWs in the lower frequency range. Their instrument sensitivity goals all reach $\Omega_{gw} = 10^{-17}$ in terms of critical density. The main issue is the level of foreground and whether foreground could be separated.

When both CMB polarization experiment and space GW detectors detect traces of inflationary GWs, important information in the inflation and after inflation (e.g., the reheating processes; Nakayama et al., 2008a, 2008b) can be probed.

## 8. Testing General Relativity and Studying Black Hole Physics with Precision in Strong Gravity

An important scientific goal of LISA (LISA Study Team, 2000) is to test general relativity and to study black hole physics with precision in strong gravity. With better precision in 100 nHz-1 mHz frequency range, ASTROD is going to push this goal further in many aspects. These include testing strong-field gravity, precision probing of Kerr spacetime and measuring/constraining the mass of graviton. Further studies in these respects would be of great value.

## 9. Outlook

The GW detection in the 100 nHz – 1 mHz band is important to the study of black hole co-evolution with galaxies, the determination of the equation of state of dark energy, testing relativistic gravity and probing the inflationary physics. With current technology development, we are in a position to explore deeper into the origin of gravitation and our universe. The current and coming generations are holding such promises.

## Acknowledgements

I would like to thank Swara Ravindranath, Varun Sahni and Jun'ichi Yokoyama for helpful discussions. I would also like to thank the referees for critical comments which helped to improve the manuscript. This work is supported in part by the National Science Council (under Grant No. NSC101-2112-M-007-007) and the National Center of Theoretical Sciences (NCTS).